\documentclass{ws-procs9x6}
\def\etal {{\it et al.}}

\usepackage{epstopdf}

\begin{document}
\title{LASER TIME-TRANSFER AND SPACE-TIME REFERENCE IN ORBIT}

\author{P.\ BERCEAU and L.\ HOLLBERG$^*$}

\address{Department of Physics and Hansen Experimental Physics Laboratory\\ 
Stanford University, Stanford, CA 94305-4085\\
$^*$E-mail: leoh@stanford.edu}

\begin{abstract}
A high performance Space-Time Reference in orbit could be realized using a stable atomic clock in a precisely defined orbit and linking that to high accuracy atomic clocks on the ground using a laser based time-transfer link.  This would enhance performance of existing systems and provide unique capabilities in navigation, precise timing, earth sciences, geodesy and the same approach could provide a platform for testing fundamental physics in space. Precise laser time- and frequency-transfer from the ground to an orbiting satellite would make it possible to improve upon the current state of the art in timing (about 1 to 30\,ns achieved with GPS) by roughly a factor of 1000 to the 1\,ps level.  
\end{abstract}

\bodymatter

\section{Motivation}

There have been tremendous advances in the performance of atomic frequency standards (clocks) over the past 40 years, and, for compelling reasons, there are growing efforts to put more advanced atomic clocks into space.   Prominent examples are the PHARAO cold-cesium atomic clock that is part of the European ACES mission\cite{ESA_site} scheduled to fly on the International Space Station about 2016, the compact Hg$^+$ ion standard of JPL designed for space applications, and other promising systems under development for the future (the ESA Space Optical Clock,\cite{Schiller_SOC_project} the DARPA Slow Beam Optical Clock\cite{DARPA_AOSense}). 
Advanced laser systems have already vastly improved the performance of atomic clocks and optical frequency synthesis and division. Lasers can do the same for time transfer to space.

\section{System concept}

\begin{figure}[t]
\begin{center}
\psfig{file=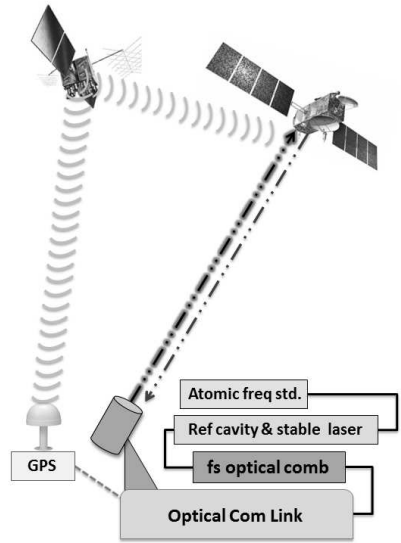,width=0.5\textwidth}
\end{center}
\caption{Diagram of a high performance optical time/frequency transfer link between the ground and an STR satellite. Also shown is a GPS satellite in a higher orbit. GPS signals detected on the STR satellite provide information about the satellite position and orbit, while the STR measures the GPS clocks relative to the ground clock with high precision and from above the troposphere and ionosphere.}
\label{OpticalLink}
\end{figure}

The basic system concept is to take advantage of the very high stability and high accuracy available from cold-atom atomic clocks, as well as the precise timing and optical frequency division provided by femotosecond  lasers, and then leverage instrumentation and technologies developed for other applications in space to achieve a robust high precision time transfer system.   The other technologies required include: laser communication (LaserCom) links between ground and space,\cite{LaserCom} ultrafast electronics and photodetectors, and robust fiber-optic telecom technologies.  LaserCom links to the satellite can transport the high performance atomic time from state-of-the-art ground clocks to the satellite. Integrating these systems and technologies would allow us to create a very high performance inertial Space-Time Reference (STR) satellite with unprecedented performance, that would serve as a stable and accurate reference for coordinate position and time.
A block diagram of the major system components 
is shown in Fig.\ \ref{OpticalLink}. 

 The precise time and frequency reference on the ground comes from a high performance atomic frequency standard (“clock”) as found at a national standards laboratory.  A stable CW laser (or microwave source) is locked to the narrow atomic transition; the stable CW laser, in turn, stabilizes the self-referenced fs-optical frequency comb.  The stabilized fs-optical comb is a train of optical pulses that are phase-coherent with the optical comb and the stable CW laser. The train of ultra-short optical pulses serves as the heart of the optical time transfer system.  The precise timing information available from the comb and optical pulses are then transferred to a two-way LaserCom optical link to the satellite. The individual sub-systems, depicted by boxes in Fig.\ \ref{OpticalLink}, have been developed in national standards laboratories and universities and have demonstrated outstanding performance for different specific applications. Many critical scientific questions and technical challenges must be addressed, including: optimal encoding of precise timing and time (epoch) information on the optical pulses, pointing and satellite tracking, signal acquisition, mitigation of atmospheric turbulence and group delay effects and dispersion, determination of time (epoch) relative to GNSS time and coordinate systems and orbit determination.

\section{Goals expected and outcome}

Our current studies are investigating how to optimally transfer precise atomic time (epoch), time-interval, and frequency from high accuracy atomic clocks on the ground to satellites in space (or alternatively between satellites in space).  We aim to achieve the following:
%\indent --\, $\leq$1 ps timing precision between high accuracy atomic frequency standards on the ground and the stable clock in %orbit,\\
%\indent --\, the atomic clock on the STR should support timing stability at the 1\,ps level for time intervals between updates %from ground clocks,\\
%\indent --\, the precision laser Time-transfer and with a orbit above the atmosphere and most of the ionosphere the STR %satellite will provide higher precision measurements of GNSS clocks in orbit,\\
%\indent --\, sub-mm orbit determination from a combination of the two-way LaserCom system and signals from a multitude of GNSS %satellites, and perhaps augmented by other systems (two-way microwave link, VLBI),\\ 
%\indent --\, high precision Time- and frequency-transfer, enabling comparison of ground clocks at the $10^{-17}$ fractional %frequency.  
(i) $\leq$1 ps timing precision between high accuracy atomic frequency standards on the ground and the stable clock in orbit,
(ii) the atomic clock on the STR should support timing stability at the 1\,ps level for time intervals between updates from ground clocks,
(iii) the precision laser time-transfer and with an orbit above the atmosphere and most of the ionosphere the STR satellite will provide higher precision measurements of GNSS clocks in orbit,
(iv) sub-mm orbit determination from a combination of the two-way LaserCom system and signals from a multitude of GNSS satellites, and perhaps augmented by other systems (two-way microwave link, VLBI), and 
(v) high precision time- and frequency-transfer, enabling comparison of ground clocks at the $10^{-17}$ fractional frequency.  
With the right combination of existing clocks, lasers and related technologies it is now feasible to improve the timing precision in space and time distribution around the world by roughly a factor of a thousand, from the few nanosecond from GNSS systems to the picosecond level.  In the laboratory we have been evaluating system configurations, architectures and components that can give high performance in timing, time and frequency and that are robust and of sufficiently low power that it is reasonable to expect that these could be used in space in the near future.  Our optical timing and time-transfer system is based on a self-referenced fs mode-locked Erbium fiber laser (Menlo Systems) that can be locked to either an RF or optical frequency reference. Currently we simulate the time transfer link using a 4\,km round-trip optical fiber between buildings on campus.  Even without fiber link stabilization we can achieve timing noise densities of much less than 1\,ps on the optical pulses and timing noise densities of several fs/$\surd \mathrm{Hz}$ on the optical carrier.    

Dedicated optical links on the ground have already demonstrated timing precision at the femtosecond level over fiber and short free-space distances, so the stability of the laser systems and ground clocks would not be the limiting factor.\cite{Bluhm_PRL}  Achieving ps timing precision between ground and space would be a major advance for satellite systems and could significantly improve the performance of existing space navigation systems.

This project is motivated by trying to bring the capabilities of advanced atomic clocks to space.  The widespread practical utility of precise time and space reference naturally bring potentially synergistic connections to existing and growing GNSS and applications, including proposed geodetic and geoscience missions where precise orbit determination and relative position measurements are critical for validating terrestrial reference frames, understanding earth plate motions, water/ice movement, and geodetic and earth science missions. This type of approach could serve as an exceptional platform for testing aspects of spacetime dependence of fundamental physics including General Relativity and searches for new physics beyond the Standard Model. For some fundamental physics experiments and earth science applications it could be advantageous to operate the STR satellite as a drag-free satellite using an inertial test mass reference to further reduce external perturbations to the orbit.\cite{Private_Com}

\section*{Acknowledgments}
We gratefully acknowledge support from the NASA Fundamental Physics and DARPA-QUASAR programs.

\end{document}